\documentclass[twocolumn]{aastex701}

\usepackage{enumitem}

\usepackage{newtxtext,newtxmath}
\usepackage{bm,ctable,verbatim,soul}
\usepackage{upgreek}
\usepackage{amsmath,cancel}
\usepackage{sankey}
\usepackage{subcaption}
\newcommand{\datastatement}[1]{\begin{small}\section*{Data Availability Statement}\end{small}{\noindent #1}\vspace{5pt}}

\definecolor{darkgreen}{RGB}{0,100,0}
\definecolor{mlblue}{RGB}{153,153,255}
\newcommand{\edit}[1]{#1}

\newcommand{\ml}[1]{}

\newcommand{\unit}[1]{\ensuremath{\,\mathrm{#1}}}
\newcommand{\Msun}{\unit{M_\odot}}

\newcommand{\GIZMO}{{\sc GIZMO }}
\newcommand{\MUSIC}{{\sc MUSIC }}
\newcommand{\AHF}{{\sc AHF }}

\newcommand{\figref}[1]{Fig.~\ref{#1}}
\newcommand{\myeqref}[1]{Eq.~\eqref{#1}}
\newcommand{\secref}[1]{\S~\ref{#1}}

\graphicspath{{figure/}}

\begin{document}
\title{Enhanced Multiphase Circumgalactic Medium and Gas Cycling in Galaxy Mergers}
\author[0009-0007-9645-4793]{Maolan Yang}
\affiliation{Shanghai Astronomical Observatory, Chinese Academy of Sciences, Shanghai  200030, P.R. China}
\affiliation{University of Chinese Academy of Sciences, 19A Yuquan Road, Beijing 100049, P.R. China}
\email{yangmaolan@shao.ac.cn}

\author[0000-0001-9658-0588]{Suoqing Ji}
\affiliation{Center for Astronomy and Astrophysics and Department of Physics, Fudan University, Shanghai 200438, P.R. China; sqji@fudan.edu.cn; fyuan@fudan.edu.cn}
\affiliation{Key Laboratory of Nuclear Physics and Ion-Beam Application (MOE), Fudan University, Shanghai 200433, P.R. China}
\email{sqji@fudan.edu.cn}

\author[0000-0002-1109-1919]{Robert Feldmann}
\affiliation{Department of Astrophysics, University of Zurich, Zurich CH-8057, Switzerland; robert.feldmann@uzh.ch}
\email{robert.feldmann@uzh.ch}

\author[0000-0003-3564-6437]{Feng Yuan}
\affiliation{Center for Astronomy and Astrophysics and Department of Physics, Fudan University, Shanghai 200438, P.R. China; sqji@fudan.edu.cn; fyuan@fudan.edu.cn}
\email{fyuan@fudan.edu.cn}

\author[0000-0002-3430-3232]{Jorge Moreno}
\affiliation{Department of Physics and Astronomy, Pomona College, Claremont, CA 91711, USA}
\affiliation{Carnegie Observatories, 813 Santa Barbara St., Pasadena, CA 91101, USA}
\email{jorge.moreno@pomona.edu}

\author[0000-0002-2853-3808]{Taotao Fang}
\affiliation{Department of Astronomy, Xiamen University, Xiamen, Fujian 361005, People's Republic of China}
\email{fangt@xmu.edu.cn}

\author[0000-0002-2651-7281]{Coral Wheeler}
\affiliation{Department of Physics and Astronomy, California State Polytechnic University, Pomona, Pomona, CA 91768, USA}
\email{cwheeler@cpp.edu}

\author[0000-0002-6864-7762]{Luigi Bassini}
\affiliation{Department of Astrophysics, University of Zurich, Winterthurerstrasse 190, Zurich CH-8057, Switzerland}
\email{gigibassini@gmail.com}

\author[0000-0002-6593-8820]{Jing Wang}
\affiliation{Kavli Institute for Astronomy and Astrophysics, Peking University, Beijing 100871, People's Republic of China}
\email{jwang_astro@pku.edu.cn}

\author[0000-0002-7541-9565]{Jonathan Stern}
\affiliation{School of Physics and Astronomy, Tel Aviv University, Tel Aviv 69978, Israel}
\email{sternjon@tauex.tau.ac.il}

\author[0000-0002-4900-6628]{Claude-Andr\'e Faucher-Gigu\`ere}
\affiliation{CIERA and Department of Physics and Astronomy, Northwestern University, 1800 Sherman Ave, Evanston, IL 60201, USA}
\email{cgiguere@northwestern.edu}

\author[0000-0002-1666-7067]{Du\v{s}an Kere\v{s}}
\affiliation{Center for Astrophysics and Space Sciences, University of California San Diego, San Diego, CA 92093, USA}
\email{dkeres@physics.ucsd.edu}

\begin{abstract}
We investigate the impact of galaxy mergers on the circumgalactic medium (CGM) using the FIREbox cosmological hydrodynamic simulation. By comparing matched samples of merging and isolated galaxies with stellar masses $M_\star \sim 10^{10}$--$10^{11} \, \mathrm{M}_{\odot}$ at $z=0$ and merger mass ratios larger than $1:10$, we find that mergers significantly alter CGM properties. Merging systems exhibit enhanced radiative cooling, leading to shorter cooling times than free-fall times across large CGM volumes. This results in amplified multiphase structure and increased cool/cold gas content ($T \sim 10^4\,\mathrm{K}$) compared to isolated galaxies. Both inflow and outflow mass fluxes are elevated across a wide range, and the enhancement is mostly pronounced in cool ($\sim10^4\unit{K}$) and hot ($\sim10^6\unit{K}$) phases by at least $\sim$1 dex in mergers, with cool gas primarily generated in-situ via radiative cooling rather than from pre-existing streams. Gas cycling analysis reveals that mergers fundamentally accelerate CGM processing, amplifying the effective transfer rate from cold/cool cosmic inflow to galaxy inflow by factors of \edit{$\sim 27$} through rapid cycling of inflowing gas through intermediate CGM phases, efficiently fueling the ISM and star formation. The enhanced cool gas content in mergers produces elevated column densities for low- and intermediate-temperature ion species in the inner CGM, while high-temperature species remain largely unaffected.
\end{abstract}


\section{Introduction}
\label{sec:intro}

The circumgalactic medium (CGM) is a crucial component in galaxy formation and evolution. Although its strict definition is not universally agreed upon in the community, the CGM is generally considered to be the gas residing in the region between the edge of a galaxy and the virial radius of its host halo. In most $L^\star$ galaxies, the CGM is thought to be dominated by hot, diffuse gas with a temperature of $\sim10^6\,\mathrm{K}$ \citep{Munch1961,Bahcall1969}, while numerous observations reveal the widespread presence of cooler components, leading to a multiphase CGM scenario \citep{Tumlinson2017,Prochaska2017b,Zahedy2019,Werk13,Zhu2013,Qu2022}. The properties of the CGM are shaped by various large-scale processes and small-scale physical mechanisms \citep{FG23}. The former include the baryon cycle \citep{Wright2024,Angles-Alcazar2017,Donahue2022}, in which gas is accreted from the intergalactic medium and flows out from galaxies via stellar feedback and AGN feedback, and virial heating by the gravitational potential of the host halo \citep{Stern2020}. These are further regulated by various small-scale physical mechanisms such as radiative cooling \citep{McCourt2012,Voit2017,Gronke2018}, turbulent mixing layers \citep{Ji2019,Fielding2020,Tan2021}, magnetic fields \citep{Ji2018,Gronke2020,Nelson2020,Voort2021}, and cosmic rays \citep{Ji2020,Ruszkowski2017,Hopkins2020,Ji2021}. Due to the large amount of baryons contained in the CGM, it plays a significant role in galaxy formation and evolution by acting as a reservoir of gas and a conduit connecting galaxies to their surrounding environment, thereby regulating the star formation process within galaxies.

Galaxy mergers have been extensively studied from both observational and theoretical perspectives. Observationally, mergers are believed to play a key role in triggering star formation \citep{Ellison2008,Patton2013,Huang2025}, galaxy quenching \citep{Ellison2022}, active galactic nucleus (AGN) activity \citep{Ellison2011,Liu2011,Jin2021}, and affecting the interstellar medium (ISM) by, for example, decreasing metallicity and supplying cold gas \citep{Ellison2008}. Theoretically, the violent perturbations of the gravitational potential and the transfer of angular momentum during mergers are thought to drive these phenomena. Modern numerical simulations, both idealized and cosmological, now allow us to investigate the detailed physical processes occurring during mergers and to study the properties of the galaxies involved with greater sophistication \citep[e.g.,][]{Moreno2015,Moreno2019,Patton2020,Moreno2021,Sparre2022,Deng2025}.

In particular, modern cosmological simulations and zoom-in simulations provide a powerful tool to study mergers in a realistic context, revealing the interplay between mergers and their environments. During a merger, gas inflows from the CGM are enhanced by the interaction between the two galaxies \citep{Sparre2022}, leading to increased star formation rates \citep{Patton2020} and potentially triggering AGN activity \citep{Quai2023}. The CGM itself is also affected by the merger, which can store the expelled gas from the infalling satellite galaxy via tidal forces, stellar feedback and/or AGN feedback \citep{Hani2018}. These processes may shape the observed properties of the CGM, including its metal content, multiphase structure, spatial distribution, and kinematic state. Understanding this two-way interaction between mergers and the CGM is therefore crucial for a complete picture of how mergers drive galaxy evolution, yet the detailed physical mechanisms remain incompletely understood.

Despite significant progress in both simulations and observations, comprehensive studies of the CGM during merger events remain limited. While pioneering simulation work has begun to explore CGM-merger connections \citep[e.g.,][]{Hani2018,Sparre2022,Carr2025}, most studies have employed idealized merger setups in isolation, which necessarily simplify the cosmological environment and the continuous baryon cycling between the intergalactic medium, CGM, and galaxies. Furthermore, the primary focus has typically been on galactic properties such as star formation and ISM kinematics \citep{Moreno2019,Patton2020,Moreno2021}, with the CGM often characterized through integrated properties rather than a detailed examination of its multiphase structure, spatial evolution, and the interplay of multiple physical processes (e.g., cooling, mixing, feedback) acting simultaneously. Observationally, while absorption line and radio emission studies have revealed rich CGM structures around galaxies \citep{Tumlinson2017,Prochaska2017b,Wang2023}, systematically connecting these observations to merger-driven processes remains challenging. Consequently, key questions regarding the response of the multiphase CGM to mergers and the underlying physical mechanisms persist, addressing which requires simulations that combine cosmological environments with CGM multiphase structure and track the relevant physical processes. 

\begin{figure*}
    \centering
    \includegraphics[width=0.85\linewidth]{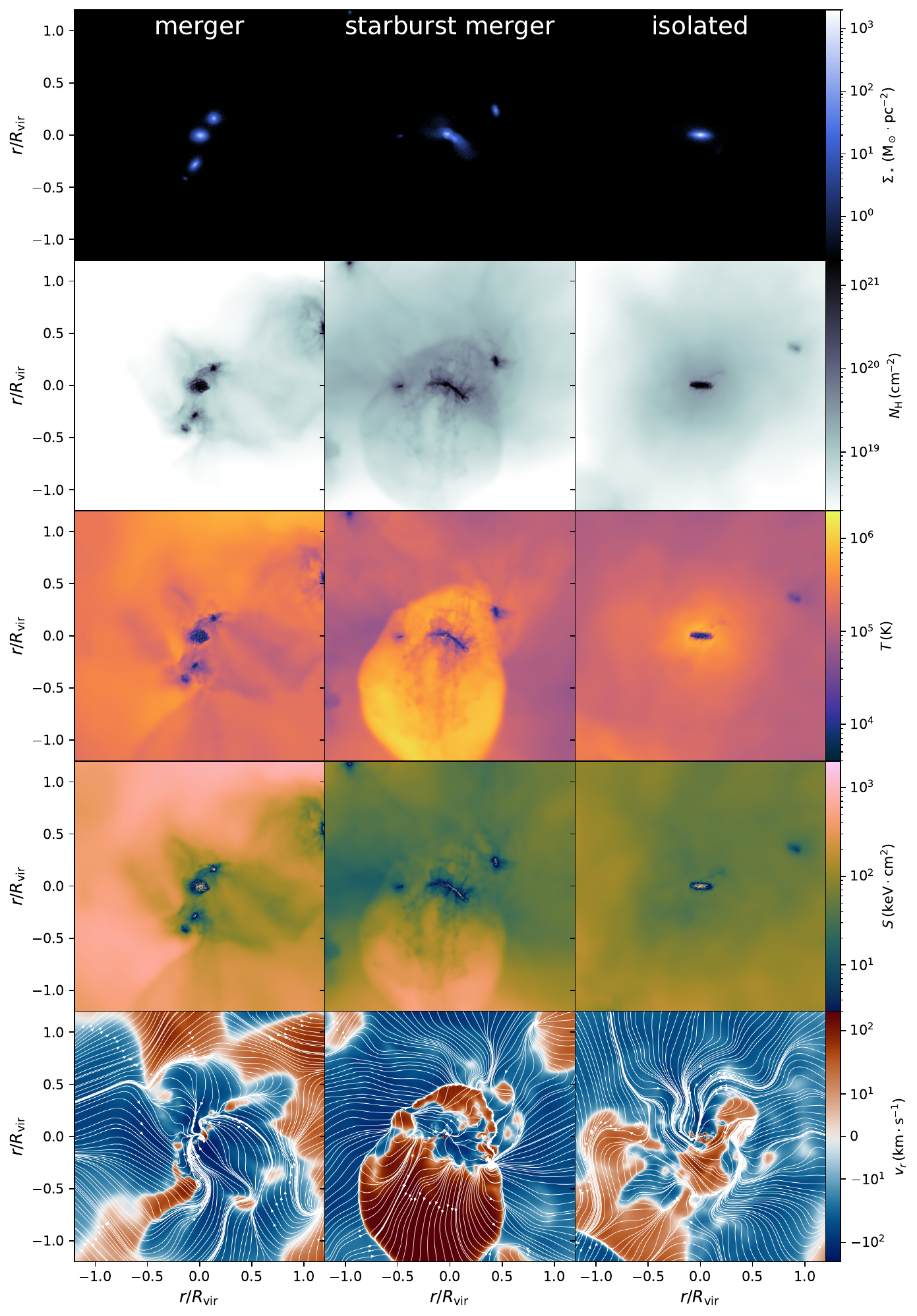}
    \caption{Projected distributions of key physical properties across three galactic systems. From top to bottom: stellar surface density, hydrogen column density, gas density-weighted temperature, entropy, and radial velocity. The panels show a typical galaxy merger hosting three galaxies with stellar masses of $2.2\times10^{10}\unit{M_\odot}$, $8.4\times10^{9}\unit{M_\odot}$ and $8.0\times10^{9}\unit{M_\odot}$ (left column), a starburst merger with stellar mass of $2.1\times10^{10}\unit{M_\odot}$ (or $6.5\times10^{9}\Msun$ and $7.7\times10^9\Msun$ when the two galaxies start merging at $z=0.12$) (middle column), and a corresponding isolated control galaxy with stellar mass of $4.1\times10^{10}\unit{M_\odot}$ (right column) at $z=0$. The merger systems exhibit more disturbed CGM structures than the isolated galaxy, with more cool gas and higher inflow velocities. The starburst merger shows particularly strong outflow velocities due to intense stellar feedback triggered by the interaction.}
    \label{fig:spatial_properties}
\end{figure*}

In this work, we address these questions by conducting a systematic study of the CGM during galaxy mergers within a fully cosmological framework. We utilize the modern hydrodynamic cosmological simulation FIREbox \citep{Feldmann2023}, which is based on the FIRE-2 physics model\footnote{\href{http://fire.northwestern.edu}{http://fire.northwestern.edu}} \citep{Hopkins2018} and combines a large cosmological volume with the key physical processes governing the evolution of the multiphase CGM. This allows us to investigate mergers in their proper cosmological context while resolving the detailed interplay between mergers and the multiphase CGM. As an illustration, \figref{fig:spatial_properties} shows two examples of galaxy mergers (a triple system and a starburst merger) alongside an isolated control galaxy from FIREbox at $z=0$. The merging system on the left-most column hosts three galaxies with stellar masses of $2.2\times10^{10}\unit{M_\odot}$, $8.4\times10^{9}\unit{M_\odot}$ and $8.0\times10^{9}\unit{M_\odot}$, the middle system has a total stellar mass of $2.1\times10^{10}\unit{M_\odot}$ (or $6.5\times10^{9}\Msun$ and $7.7\times10^9\Msun$ when the two galaxies start merging at $z=0.12$), while the isolated galaxy on the right-most panel has a stellar mass of $4.1\times10^{10}\unit{M_\odot}$. The merging galaxies exhibit notably more disturbed CGM structures compared to the isolated galaxy, with a higher prevalence of cool gas and increased inflow velocities. The starburst merger, in particular, displays pronounced outflow velocities driven by intense stellar feedback triggered by the interaction. We detail our methods and present comprehensive results in the following sections.

The paper is organized as follows. In \S\ref{sec:methods}, we describe the numerical simulations with the physics included in our study. In \S\ref{sec:results}, we show the results of our simulations. We finally discuss the implications of our results and conclude in \S\ref{sec:discussion}.

\section{Methods} \label{sec:methods}
\subsection{An overview of the FIREbox simulation}

This study utilizes the FIREbox cosmological hydrodynamic simulation \citep{Feldmann2023}, which implements the FIRE-2 physics framework \citep{Hopkins2018} using the meshless finite-mass methodology incorporated in the \GIZMO code \citep{Hopkins2015}. The FIREbox simulation suite models the co-evolution of gas, stellar populations, and dark matter within a periodic cosmological volume of $(15 \, \mathrm{Mpc}/h)^3$. Our analysis employs the highest-resolution simulation from this suite, designated ``FB1024'' in \citet{Feldmann2023}, which incorporates $1024^3$ gas particles and $1024^3$ dark matter particles. This configuration achieves a mass resolution of $m_\mathrm{b} = 6.26 \times 10^4 \, \mathrm{M}_{\odot}$ for baryonic matter and $m_\mathrm{DM} = 3.35 \times 10^5 \, \mathrm{M}_{\odot}$ for dark matter. The simulation adopts cosmological parameters consistent with Planck Collaboration measurements \citep{Planck2016}: $h = 0.6774$, $\Omega_\mathrm{m} = 0.3089$, $\Omega_\Lambda = 0.6911$, $\Omega_\mathrm{b} = 0.0486$, $\sigma_8 = 0.8159$, and $n_\mathrm{s} = 0.9667$. Initial conditions were generated at redshift $z = 120$ using the \MUSIC (Multi-Scale Initial Conditions) code \citep{Hahn2011}.

FIREbox incorporates the comprehensive FIRE-2 physics model, which accounts for critical baryonic processes including radiative cooling, heating mechanisms, star formation, and stellar feedback. The thermal evolution of gas spans a wide temperature range from $10\,\mathrm{K}$ to $10^{10}\,\mathrm{K}$, with cooling and heating calculations that incorporate free-free, photoionization, recombination, Compton, photoelectric, and dust-coupling processes. The simulation accounts for both local radiation sources and the metagalactic ultraviolet background as characterized by \citet{FG09}. Star formation occurs exclusively in gas that simultaneously satisfies multiple physical criteria: self-gravitating conditions, number densities exceeding $300\,\mathrm{cm}^{-3}$, Jeans instability, and molecular (self-shielding) state. Stellar feedback is implemented through multiple channels, including Type Ia and Type II supernovae that inject mass, momentum, energy, and metals into the surrounding medium, as well as mass loss from OB and AGB stars via stellar winds. Additionally, the simulation incorporates radiative feedback mechanisms from these stellar populations. It should be noted that the FIREbox simulation we analyze in this paper does not include active galactic nucleus (AGN) feedback or cosmic ray physics.

\subsection{Sample galaxy selection}\label{sec:sample_selection}

To investigate the effects of galaxy mergers on CGM properties, we constructed statistically meaningful samples of both merging and isolated galaxies from the FIREbox simulation at $z=0$. Our selection procedure followed a hierarchical approach, beginning with host halo identification and concluding with matched merger-control galaxy pairs.

First, we utilized the halo catalogs described in \cite{Feldmann2023}, which were generated by \AHF \citep{AHF0,AHF1}, to identify suitable host halos within the FIREbox data. We established a lower central galaxy baryon mass threshold of $10^{10}\Msun$, which guarantees that the CGM in our sample halos is resolved by a few $\sim 10^4$ particles for the given baryon mass resolution, providing adequate sampling of CGM structure and multiphase properties. Some galaxies reside in subhalos within the virial radius of a more massive host halo, where the CGM can be affected by both the subhalo and the host halo. Since these contributions are difficult to disentangle, we excluded all subhalos identified by \AHF. After applying these criteria, we identified 70 host halos at $z=0$ suitable for further analysis.

Within these selected halos, we then classify galaxies as either merging or isolated based on their interaction history. We define a galaxy as a merging galaxy if it is either currently undergoing a merger or has experienced a merging event within the past $1\unit{Gyr}$; otherwise, it was classified as isolated. Due to the small number of host halos remaining after the previous selection, we identify merging galaxies by visual inspection. We further require a minimum merger mass ratio of 1:10, excluding interactions with much smaller satellites. We adopt a $1\,\mathrm{Gyr}$ threshold to capture the timescale over which merger-induced perturbations propagate through and relax within the CGM, corresponding to the longest characteristic timescale governing the CGM response to mergers. One physically motivated estimate is the sound-crossing time, $t_\mathrm{sc}\sim R/c_s$, where $R$ is a characteristic CGM scale and $c_s$ is the sound speed. Adopting $R\sim R_{\rm vir}\propto M_\mathrm{halo}^{1/3}(1+z)^{-1}$ and $c_s\propto T_\mathrm{CGM}^{1/2}$, this yields the scaling
\begin{align}
t_\mathrm{sc} \sim 0.6\,\unit{Gyr}\,\left(\frac{M_{\rm h}}{10^{11}\,M_\odot}\right)^{1/3} \left(\frac{T_\mathrm{CGM}}{10^{6}\,{\rm K}}\right)^{-1/2} \left(\frac{1+z}{1}\right)^{-1}.
\end{align}
If the CGM temperature tracks the virial temperature, $T_\mathrm{CGM}\sim T_\mathrm{vir}\propto M_\mathrm{halo}^{2/3}(1+z)$, the mass dependence largely cancels, yielding characteristic sound-crossing times of order $1\unit{Gyr}$ at $z\sim 0$ for our sample, justifying our choice of merger timescale. Additional relevant timescales include the dynamical time and the cooling time of the CGM, but these are generally comparable to or shorter than the sound-crossing time for our sample. This threshold is somewhat shorter than merger definitions adopted in some observational studies (e.g., \citealt{Hani2020}), but is well-suited for capturing merger-induced CGM perturbations in our simulation analysis.

For each identified merging galaxy, we construct control samples from the isolated galaxy population by matching gas and stellar mass properties. An isolated galaxy is selected as a control if it exhibits similar gas and stellar mass to a merging galaxy. Each merger is required to have at least one matching control galaxy to be included in our final sample. For mergers with multiple potential controls, we selected up to three isolated galaxies with the closest matching gas and stellar mass. To maximize statistical power, we allow individual isolated galaxies to serve as controls for multiple mergers when appropriate. This matching procedure yields a final sample of 18 merging galaxies paired with 20 control galaxies and 44 merger-control pairs for comparative analysis. There are 11 major mergers with mass ratios larger than $1:3$ and 7 minor mergers with mass ratios smaller than $1:3$ in our merging-galaxy sample.

\subsection{Properties of the selected galaxy sample}

\begin{figure}
    \centering
    \includegraphics[width=\linewidth]{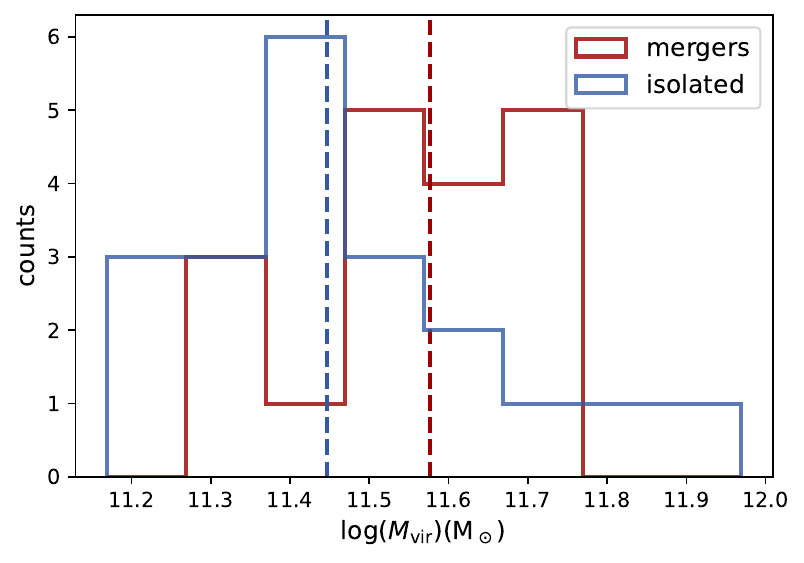}
    \caption{Virial mass distributions of host halos, showing the counts of halos hosting merging galaxies (red histogram) and halos hosting isolated galaxies (blue histogram) within given virial mass bins. The red (blue) dashed vertical lines indicate the median virial masses of merging (isolated) galaxy host halos. Virial masses of selected halos range from $10^{11}\Msun$ to $10^{12}\Msun$.}
    \label{fig:halo_virial_mass}
\end{figure}
\begin{figure}
    \centering
    \includegraphics[width=\linewidth]{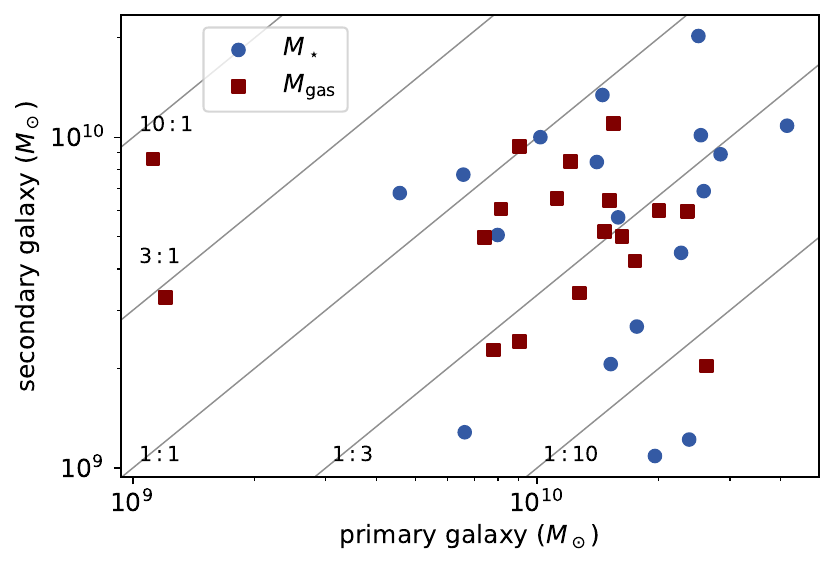}
    \caption{Stellar mass and gas mass distributions of the primary and secondary galaxies in merging pairs (blue dots and red squares, respectively). Background grey lines denote the mass ratio of the secondary galaxy to the primary galaxy, with the ratio of secondary galaxy to primary galaxy labeled near each line. Our sample covers a stellar mass range of $3\times10^9\unit{M_\odot}$ to $4\times10^{10}\unit{M_\odot}$, and a gas mass range of $5\times10^8\unit{M_\odot}$ to $2\times10^{10}\unit{M_\odot}$}.
    \label{fig:galaxy_mass}
\end{figure}
Fig.~\ref{fig:halo_virial_mass} presents the virial mass distributions of host halos for both merging (red histogram) and control (blue histogram) galaxies. The vertical dashed lines indicate the median virial masses for each population. Merging galaxies preferentially reside in slightly more massive halos than isolated systems, though the two distributions substantially overlap, indicating that halo mass alone does not strongly discriminate between the two populations. \figref{fig:galaxy_mass} shows the stellar and gas mass distributions of primary and secondary galaxies in our merger sample, displayed as blue dots and red squares, respectively. Secondary stellar and gas masses broadly correlate with primary stellar mass, as expected given our restriction to mass ratios greater than 1:10, though gas masses exhibit larger scatter at fixed stellar mass. Most galaxies in our sample are dwarf or sub-$L^\star$ systems with stellar masses ranging from $3\times10^9\unit{M_\odot}$ to $4\times10^{10}\unit{M_\odot}$, gas masses between $5\times10^8\unit{M_\odot}$ and $2\times10^{10}\unit{M_\odot}$, and host halo virial masses below $10^{11.9}\Msun$. This predominance of lower-mass systems results from both the limited volume of the FIREbox simulation and our selection criteria. While FIREbox reproduces the theoretical halo mass function within $0.2\unit{dex}$ \citep[see][Appendix A]{Feldmann2023}, the scarcity of halos with virial masses exceeding $10^{12}\Msun$ is a direct consequence of the simulation's finite volume. Additionally, our exclusion of galaxies with numerous satellites eliminates many massive halos, further limiting the available massive systems for constructing control samples.

\begin{figure}
    \centering
    \includegraphics[width=0.95\linewidth]{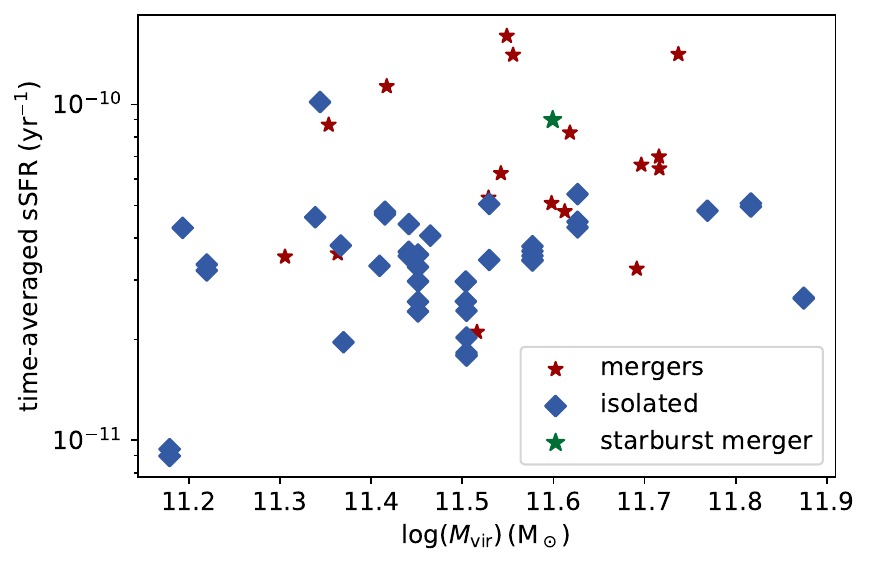}
    \caption{Time-averaged specific star formation rate ($\langle\mathrm{sSFR}\rangle$) of merging galaxies and control galaxies versus virial masses of host halos. Merging galaxies are indicated with stars while control galaxies are indicated with blue octagons. A notable merging pair with high instantaneous sSFR at $z=0$ is marked in green and labeled as ``starburst merger,'' while other merging samples are marked in red.}
    \label{fig:SFR}
\end{figure}

\figref{fig:SFR} displays the time-averaged specific star formation rates ($\langle\mathrm{sSFR}\rangle$) of merging and isolated galaxies as a function of host halo virial mass. For each merger, $\langle\mathrm{sSFR}\rangle$ is averaged from the time when the companion first enters the host halo to $z=0$. For each matched isolated control, we average over the same time interval as its corresponding merger; therefore, the same isolated galaxy can appear more than once if it is matched to mergers with different start times. \figref{fig:SFR} shows an enhancement in $\langle\mathrm{sSFR}\rangle$ between mergers and isolated galaxies, from a median value of $3.5\times10^{-11}\unit{yr^{-1}}$ in isolated galaxies to $6.5\times10^{-11}\unit{yr^{-1}}$ in mergers. Overall, this corresponds to a factor of $\sim 2$ enhancement of the time-averaged sSFR in merging galaxies relative to isolated systems. \edit{16 out of 18 mergers have a higher $\langle\mathrm{sSFR}\rangle$ than that of at least one of their corresponding control galaxies, and 35 out of 44 merger-control pairs show $\langle\mathrm{sSFR}\rangle$ enhancement}. This modest time-averaged enhancement is consistent with previous observational and simulation studies of merger-enhanced star formation \citep[e.g.,][]{Ellison2008,Patton2013,Moreno2015,Moreno2019,Moreno2021,Patton2020}. One merging pair stands out as undergoing a merger-triggered starburst phase, with instantaneous $\mathrm{sSFR}\gtrsim 5\times10^{-9}\,\mathrm{yr}^{-1}$ at $z=0$, approximately two orders of magnitude above the sample median. We refer to this system as the ``starburst merger'' in subsequent analyses and highlight it in green in \figref{fig:SFR}.

\subsection{Defining the CGM} \label{sec:define_CGM}

To isolate the CGM from the galactic ISM, we exclude all galactic regions (including the central galaxy, merging galaxies, and any satellite galaxies) from consideration. We identify galaxies by applying a friends-of-friends algorithm to stellar particles with a $\sim1.5\unit{kpc}$ (or $1\,\mathrm{kpc}/h$, which equals one code length at $z=0$ in the FIREbox simulation) linking length \citep{FOF}. Only stellar aggregations with mass exceeding $10^{-3}$ of the halo's total stellar mass are classified as galaxies. Each galaxy's region is defined as a sphere centered on its mass-weighted center with radius twice the distance from the center to the farthest stellar particle. We checked this radius selection method for every galaxy and found that it was not affected by extreme cases, such as distant stellar particles that could yield a galaxy radius that separates the galaxy and CGM poorly. We also tested other radius definitions, such as the radius of the 99th percentile stellar particle or the half-mass radius. While in some cases (for example, in isolated galaxies) these definitions work well and give results consistent with the method we adopt, they may generate too large or too small radii in some other galaxies, especially in merging ones, which may be the result of the sparse stellar distribution due to tidal effects. In this manner we identify all galaxies within each host halo and treat gas inside these galactic spheres as ISM. The outer boundary of the CGM is somewhat arbitrary; a common convention is to define the CGM as the gaseous halo within the virial radius \citep[e.g.,][]{Tumlinson2017,Hafen2019,Sparre2020}. In this work, we therefore define the CGM as all gas within the host halo's virial radius but outside any galactic spheres. This method yields galaxy radii of $15$--$55\unit{kpc}$ with a median radius of $21\unit{kpc}$, or galaxy-to-virial-radius ratios of $0.08$--$0.30$ with a median of $0.11$ for the central galaxies in our sample.

The method we use to define the CGM is not exactly the same as that used in some previous studies \citep[e.g.,][]{Sparre2020,Stern2021}. Some studies utilize several halos and use a fixed ratio to $R_\mathrm{vir}$ to define the inner boundary of the CGM, which is suitable for their specific scientific goals. However, we find some limitations of this approach in our case, including, for example, some galaxies having very extended and sparse stellar distributions, some halos hosting multiple galaxies with large mass differences, and large-scale stellar tidal structures in some merging systems. These factors can lead to contamination of CGM measurements by ISM gas if a fixed fraction of $R_\mathrm{vir}$ is used as the inner CGM boundary. Some studies use similar methods to ours to define the CGM \citep{Hafen2019}, while the specific details such as the way to identify galaxies may vary. Nevertheless, we note that these different CGM definitions --- including ours --- lead to similar CGM inner boundaries in most cases and thus should not significantly affect our main results.

\subsection{Scatter calculation} \label{sec:scatter_calculation}

Because the CGM distribution is non-axisymmetric (see \figref{fig:spatial_properties}), we introduce the following method to summarize the population-level radial trends while retaining information about the intrinsic scatter. For each galaxy we compute the radial distribution of values at each radius and derive that galaxy's median radial profile and per-radius distribution. We use the median profile of all the galaxy-level profiles to represent each population and bootstrap all the galaxy-level profiles to estimate the uncertainty of the representative profile of each population. To illustrate the full spread, we \edit{compute the distribution of values at each radius} across all galaxies in the same population and display the 16th–84th percentile range as a shaded band in the figures. Thus the bootstrapped range quantifies the uncertainty of the population median, whereas the full percentile range reflects the intrinsic galaxy-to-galaxy and intra-halo variation, including non-axisymmetric CGM structure. For the starburst merger (the sole member of its category) the plotted profile reduces to its individual profile and the shaded band to its own 16th–84th percentile range.

We apply an analogous procedure when computing mass and metal fluxes through spherical shells (\secref{sec:flows}): we compute the flux through each shell for every galaxy, then present the median flux at each radius for the merging and isolated populations. At a given radius each galaxy contributes a single flux value; the scatter is thus the 16th–84th percentile range across galaxies in the population. No scatter is shown for the single starburst merger.

\section{Results} \label{sec:results}

\subsection{Spatial properties}

Figure~\ref{fig:spatial_properties} illustrates the distinct representative CGM morphologies across different galactic environments. At $z=0$, we select one typical galaxy merger that includes three galaxies with stellar masses of $2.2\times10^{10}\unit{M_\odot}$, $8.4\times10^{9}\unit{M_\odot}$ and $8.0\times10^{9}\unit{M_\odot}$, the starburst merger highlighted in \figref{fig:SFR} with a total stellar mass of $2.1\times10^{10}\unit{M_\odot}$ (or $6.5\times10^{9}\Msun$ and $7.7\times10^9\Msun$ when the two galaxies start merging at $z=0.12$; because the two galaxies are strongly merging, it is difficult to identify their individual masses) and an isolated control galaxy with a stellar mass of $4.1\times10^{10}\unit{M_\odot}$ to illustrate this difference. \footnote{The figure also contains a few small satellite galaxies, such as the lowest one in the typical merger and the two outside the central region of the starburst merger. They are not listed explicitly because their masses are less than one-tenth of the main galaxies (see \S\ref{sec:sample_selection}).}

The typical galaxy merger (left column) exhibits a triple-galaxy system clearly visible in the stellar surface density panel, with three distinct galactic structures in various stages of interaction. This system displays asymmetric gas distributions with pronounced tidal features and enhanced gas density in the bridge regions connecting the interacting galaxies. The temperature distribution reveals significant cool gas concentrations along these interaction interfaces, accompanied by low-entropy regions indicative of enhanced radiative cooling. The velocity structure demonstrates complex patterns comprising both inflows toward the galactic centers and non-equilibrium motions in the extended halo.

The starburst merger (middle column) presents two galaxies in direct collision, as shown in the stellar surface density panel, creating more extreme conditions characterized by powerful bipolar outflows evident in the temperature, entropy, and radial velocity maps, with material being expelled at velocities exceeding $200\,\mathrm{km\,s^{-1}}$. These outflows correspond to regions of elevated entropy, signifying thermal energy deposition from vigorous stellar feedback processes triggered by the galactic collision.

In contrast, the isolated control galaxy (right column) displays a more regular, quasi-spherical distribution across all properties, exhibiting a smooth temperature gradient and ordered infall pattern characteristic of a virialized halo in hydrostatic equilibrium. Although the velocity panel shows some inhomogeneity, the magnitude is smaller than in the mergers. The entropy structure in merging systems reveals substantial mixing between different gas phases, disrupting the stratification evident in the isolated case. These visualizations illustrate how galaxy interactions fundamentally alter the thermodynamic and kinematic properties of the CGM, creating conditions conducive to enhanced gas cooling and accretion, as quantified in subsequent analyses.

\subsection{CGM gas phases}

\begin{figure}
    \centering
    \includegraphics[width=\linewidth]{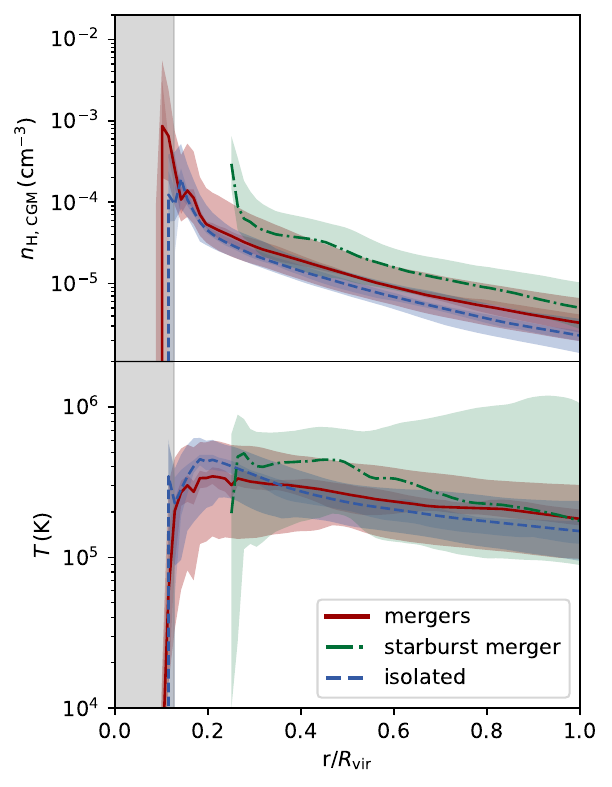}
    \caption{Radial profiles of hydrogen number density (top) and temperature (bottom) for merging systems (red solid line), the starburst merger (green dash-dotted line), and isolated galaxies (blue dashed line), with galactocentric radius normalized to each halo's virial radius $R_\mathrm{vir}$. Profiles are sampled over all galaxies in each category at $z=0$. Regions identified as galaxies are excluded from the analysis, while we use a gray band near $r/R_\mathrm{vir}=0$ to represent the median galaxy radius in our sample. The outer lighter shaded regions, colored red, green, and blue, denote the 16th to 84th percentile ranges within each corresponding sample category. The inner darker shaded region around each line shows the 16th to 84th percentiles of the median value calculated via bootstrapping. Merging galaxies exhibit marginally higher gas densities in the outer CGM and slightly lower temperatures in the inner CGM compared to isolated systems. However, the slightly higher CGM hydrogen density is the result of slightly higher halo mass in mergers, and these differences remain relatively modest, confirming that our selected samples possess comparable halo baryon content.}
    \label{fig:compare_density_and_T}
\end{figure}

While the \figref{fig:spatial_properties} illustrates specific examples of different galactic systems, here we present the statistical analysis of CGM properties across our entire sample of merging and isolated galaxies at $z = 0$. \figref{fig:compare_density_and_T} displays the radial profiles of CGM hydrogen number density ($n_\mathrm{H\,CGM}$) and temperature for both populations. We note that regions identified as galaxies (as defined in \S~\ref{sec:define_CGM}) are excluded from these measurements to focus exclusively on circumgalactic material. For individual galaxies, we first compute mean radial profiles, then take the median value across each category to represent the characteristic profile of merging and isolated populations as described in \S~\ref{sec:scatter_calculation}. 

In both merging and isolated systems, $n_\mathrm{H,\,CGM}$ exhibits a characteristic decline with increasing galactocentric distance, decreasing approximately exponentially from $\sim 10^{-3}\unit{cm^{-3}}$ in the inner halo to $\sim 10^{-5}\unit{cm^{-3}}$ near the virial radius, which is similar to previous works \citep[e.g.,][]{Blitz2000,Feldmann2013,Hafen2019}. Concurrently, the temperature profile remains relatively stable at a few $10^5\unit{K}$ throughout most of the halo, consistent with the expected virial temperature of the host dark matter halos. Merging galaxies display marginally higher $n_\mathrm{H,\,CGM}$ in the outer CGM regions and slightly reduced temperatures in the inner CGM compared to their isolated counterparts, though these differences remain modest. The starburst merger exhibits a slight overall enhancement in both $n_\mathrm{H,\,CGM}$ and temperature relative to typical mergers, reflecting the impact of powerful supernova-driven outflows and their subsequent thermalization in the halo. The apparent truncation of profiles at small radii in this particular system results from the exclusion of overlapping galactic regions during the collision, where the distinction between individual galaxies becomes ambiguous. The increase in $n_\mathrm{H}$ is a consequence of the slightly higher halo masses of mergers: mergers have higher halo masses, thus the CGM density would be higher due to higher galaxy masses. But since the difference is not significant, it can not result in the tremendous difference that will be shown in the remaining sections (\secref{sec:flows} - \secref{sec:ion}).

\begin{figure*}
    \centering
    \includegraphics[width=\linewidth]{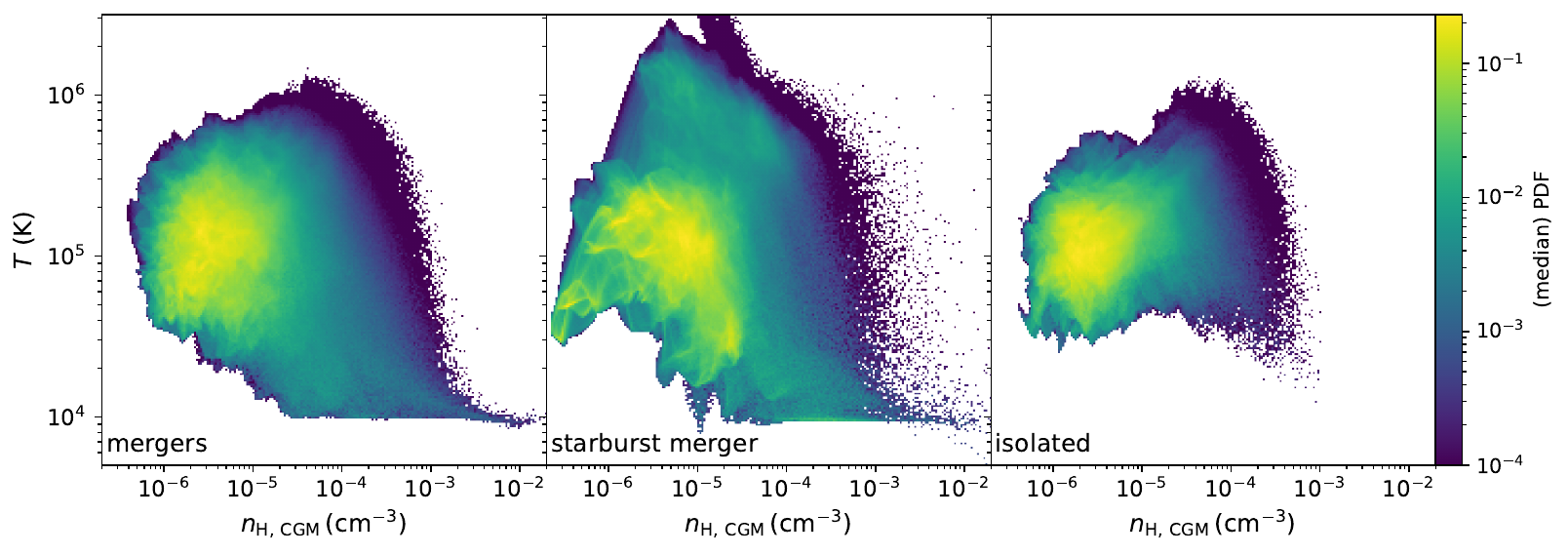}
    \caption{Density-weighted temperature-$n_\mathrm{H,\,CGM}$ phase diagram of CGM gas in merging galaxies (left), starburst mergers (middle) and isolated galaxies (right). The phase diagram is sampled over all the gas cells in all galaxies in each category at $z=0$, therefore showing a systematic difference between mergers and isolated galaxies. Color indicates the median value of probability density weighted by CGM gas mass in merging or isolated galaxies. Gas with $T\sim10^5\unit{K}$ and $n_\mathrm{H,\,CGM}\sim 10^{-6}\unit{cm^{-3}}$ dominates in both merging and isolated galaxies, but a wider distribution in phase space, especially in $10^4\sim10^5\unit{K}$, can be seen in mergers.}
    \label{fig:cgm_phase}
\end{figure*}

Although the density and temperature profiles of CGM gas in merging and isolated galaxies are broadly similar, indicating comparable total baryon content, mergers may nonetheless induce more substantial changes in the thermal structure of the CGM. To examine this, we present the density-weighted distribution of CGM gas in temperature-$n_\mathrm{H,\,CGM}$ phase space in \figref{fig:cgm_phase}, sampled over all mergers (left panel), the starburst merger (middle panel), and isolated galaxies (right panel). The color indicates the median probability density weighted by CGM gas mass for each sample. These phase diagrams are constructed by first calculating the probability density distribution of each galaxy's CGM in temperature-density space, then taking the median value across all merging or isolated systems at each phase-space location. As most CGM gas is within $10^4\unit{K} < T < 10^6\unit{K}$, we take the same notation as in \cite{Tumlinson2017} to define cold gas ($T<10^4\unit{K}$), cool gas ($10^4\unit{K}\leq T<10^5\unit{K}$), warm gas ($10^5\unit{K} \leq T < 10^{6}\unit{K}$) and hot gas ($T\geq10^{6}\unit{K}$). Although FIRE-2 models gas cooling down to $\sim10\unit{K}$ \citep{Hopkins2018}, it is less meaningful to discuss the explicit cold gas phase in the CGM because cold gas in the CGM is rare in our samples.

Compared to isolated galaxies, merging systems exhibit a substantially broader distribution in phase space, particularly in the low-temperature, high-density regime characterized by temperatures of a few $10^4\unit{K}$ and $n_\mathrm{H\,CGM}\gtrsim 3\times10^{-4}\unit{cm^{-3}}$. This extended tail toward cooler, denser conditions indicates that mergers promote the formation of cool, dense gas in the CGM, consistent with the morphological features visible in \figref{fig:spatial_properties}. The starburst merger displays an even wider distribution in phase space, with significant populations of both cool, dense gas and hot, low-density gas—the latter resulting from vigorous stellar feedback during the intense starburst phase. There is little gas with $T>10^6\unit{K}$ in either merging or isolated galaxies, as the halo mass scale considered here is $M_\mathrm{halo}\sim10^{11}\unit{M_\odot}$, which yields a virial temperature of $\sim10^5\unit{K}$. In addition, the CGM is unlikely to be fully virialized in these low-mass halos as indicated in \cite{Stern2020}.

\subsection{Enhanced multiphase inflows and outflows}\label{sec:flows}

\begin{figure*}
    \centering
    \includegraphics[width=\linewidth]{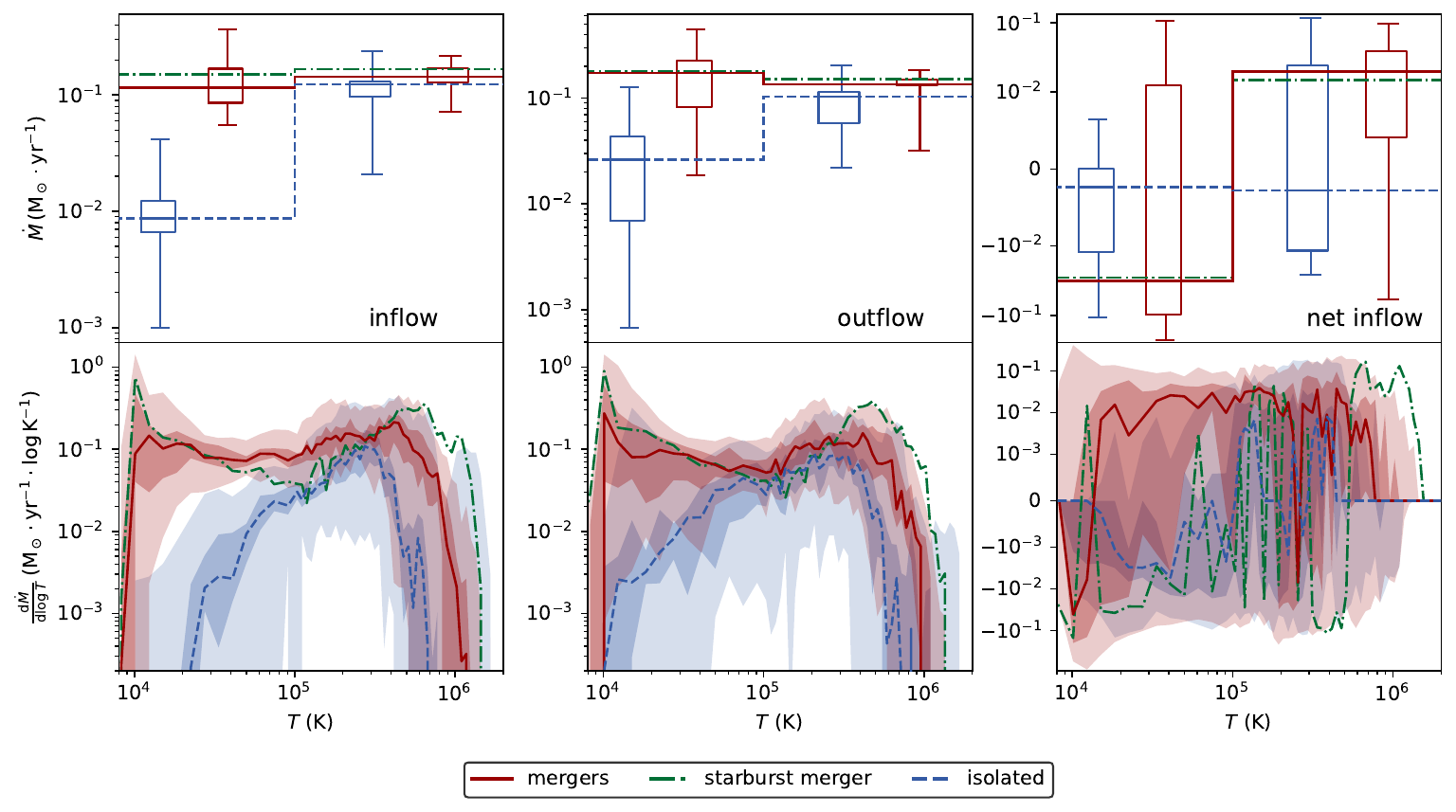}
    \caption{Mass flux of inflows (left), outflows (middle), and net inflows (right; positive values denote inflow) as a function of temperature at $R=30\unit{kpc}/h$, for merging galaxies (red solid), the starburst merger (green dash–dotted), and isolated control galaxies (blue dashed) at $z=0$. \emph{Top panels}: Cumulative mass flux integrated over $T<10^5\unit{K}$ and $T>10^5\unit{K}$. Histograms show medians; boxes indicate the 16th–84th percentile range of the bootstrapped median; whiskers give the 16th–84th percentile range across the sample. \emph{Bottom panels}: Differential mass-flux distribution, $\mathrm{d}\dot{M}/\mathrm{d}\log T$. Curves denote medians; darker (lighter) shaded regions represent the 16th–84th percentile range of the bootstrapped median (full sample). Mergers exhibit systematically higher mass fluxes than isolated galaxies, particularly for cool gas ($T\lesssim10^5\unit{K}$).}
    \label{fig:CGM_flux}
\end{figure*}

Galaxy merging events should not only impact the thermal properties of the CGM, but also the multiphase state of gas flows between the galaxies and the CGM. Here we examine how galaxy interactions affect the mass flux of CGM gas across different temperature phases. We show the cumulative mass flux integrated over $T<10^5\unit{K}$ and $T>10^5\unit{K}$ and the differential mass flux distribution ($\mathrm{d}\dot{M}/\mathrm{d}\log T$) of CGM gas at $z=0$ in \figref{fig:CGM_flux}. The mass flux is calculated at a radius of $R=30\unit{kpc}/h$ (or $\sim0.25R_\mathrm{vir}$ in our samples), which is close enough to the galaxy region we defined in \S~\ref{sec:define_CGM} in most samples, but not so small that it enters the galaxy region in a large fraction of our samples because we identify different radii for different galaxies. We also tested other radii near the galaxy and found that changing the specific value does not affect the result we show here except for the specific numerical values. This radius selection is similar to other papers that study the inflow and outflow rates in CGM \citep[e.g.,][]{Muratov2015,Pandya2021,Bassini2023}, who adopt a radius around $0.25R_\mathrm{vir}$. We also use a similar method as in the papers mentioned above to calculate mass flux. In short, we use a shell with radius between $25\unit{kpc}/h$ and $35\unit{kpc}/h$, and treat the particles falling into the halo (radial velocity $v_\mathrm{r}<0$) as inflow particles and the particles with $v_\mathrm{r}>0$ as outflow. Then we sum up the particles inside each temperature bin and normalize the sum with the temperature bin length and shell thickness. We show the cumulative mass flux in the top panels where the histograms show the median value, boxes indicate the 16th to 84th percentile range of the median value calculated via bootstrapping, and whiskers give the 16th to 84th percentile range across the sample. The differential mass flux distribution is shown in the bottom panels where the curves denote the median value and the darker (lighter) shaded regions represent the 16th to 84th percentile range of the median value calculated via bootstrapping (full sample).

In isolated galaxies (blue), \edit{ inflow and outflow mass fluxes are small and spread in a narrow temperature range. The majority of inflow and outflow is within $10^5\unit{K}$ and $10^6\unit{K}$, contributing $\sim0.13\unit{M_\odot\,yr^{-1}}$ to the total median inflow and outflow rate respectively}, while the cool gas contributes only $\sim0.007\unit{M_\odot\,yr^{-1}}$ and $\sim0.02\unit{M_\odot\,yr^{-1}}$, respectively. Because the exchange rate is modest and almost balanced in isolated galaxies, the net mass flux is small, with \edit{net outflow} of $0.002\unit{M_\odot\,yr^{-1}}$ for $T<10^5\unit{K}$ gas and $0.003\unit{M_\odot\,yr^{-1}}$ for $T>10^5\unit{K}$ gas.

The situation is markedly different in merging galaxies (red). \edit{In the bottom panel of \figref{fig:CGM_flux},} both inflow and outflow mass fluxes are substantially enhanced across all temperatures, corresponding to roughly at least 1 dex increase in $T<10^5\unit{K}$ gas exchange and modest increase in $T>10^5\unit{K}$ gas relative to isolated galaxies. \edit{Moreover, the flux of $T\sim10^{4}\unit{K}$ gas is significantly boosted, producing a pronounced peak.} As a result, unlike in isolated galaxies where warm gas dominates the inflow, merging galaxies exhibit comparable inflow rates of cool and warm gas, with median values of $\sim0.11\unit{M_\odot\,yr^{-1}}$ and $\sim0.14\unit{M_\odot\,yr^{-1}}$, respectively. The outflow also increases to median values of $\sim0.17\unit{M_\odot\,yr^{-1}}$ and $\sim0.13\unit{M_\odot\,yr^{-1}}$ for cool gas and warm gas, respectively. The net mass flux in merging galaxies is dominated by inflow and exhibits a plateau spanning from $10^4\unit{K}$ to $2\times10^{5}\unit{K}$ for the majority of mergers. The net inflow of cool gas is affected by the high outflow peak at $T\sim10^{4}\unit{K}$ and becomes negative, resulting in a net outflow with a value of $0.03\unit{M_\odot\,yr^{-1}\log K^{-1}}$, a factor of $15$ enhancement compared to isolated galaxies, but the scatter is large and most mergers show positive net inflow at higher temperatures as indicated in the lower panel. The net inflow of warm gas becomes positive compared to isolated galaxies, with a median value of $0.02\unit{M_\odot\,yr^{-1}}$, implying more efficient accretion in this phase during mergers.

For the starburst merger (green dash-dotted line), the behavior at $T\gtrsim10^{5}\unit{K}$ is similar to non-starburst mergers but with stronger hot gas exchange due to intense stellar feedback, resulting in warm-gas inflow and outflow rates of $\sim0.17\unit{M_\odot\,yr^{-1}}$ and $\sim0.15\unit{M_\odot\,yr^{-1}}$, respectively. However, the outflow at $T<4\times10^{4}\unit{K}$ is exceptionally strong and expels much of the cool gas from the galaxy. As a consequence, the inflow rate of cool gas is similar to that of non-starburst mergers, with a value of $\sim0.15\unit{M_\odot\,yr^{-1}}$, and the outflow rate increases to $\sim0.18\unit{M_\odot\,yr^{-1}}$, producing a net cool-gas outflow. The outflow strength at $T\sim10^{4}\unit{K}$ reaches the 84th percentile of the median value of our merging sample and is roughly one order of magnitude higher than the median of other mergers at the peak.

In short, the multiphase nature of gas flows is significantly amplified in mergers: while isolated galaxies exhibit gas exchange concentrated within a relatively narrow temperature range, merging systems demonstrate simultaneous strong fluxes across a broad spectrum of temperatures from cool ($\sim10^4\unit{K}$) to hot ($\gtrsim10^6\unit{K}$) phases, reflecting a fundamentally more complex and dynamic multiphase CGM structure.

\subsection{Merger-induced transition from hot-mode to cold-mode accretion}

\begin{figure*}
    \centering
    \includegraphics[width=\linewidth]{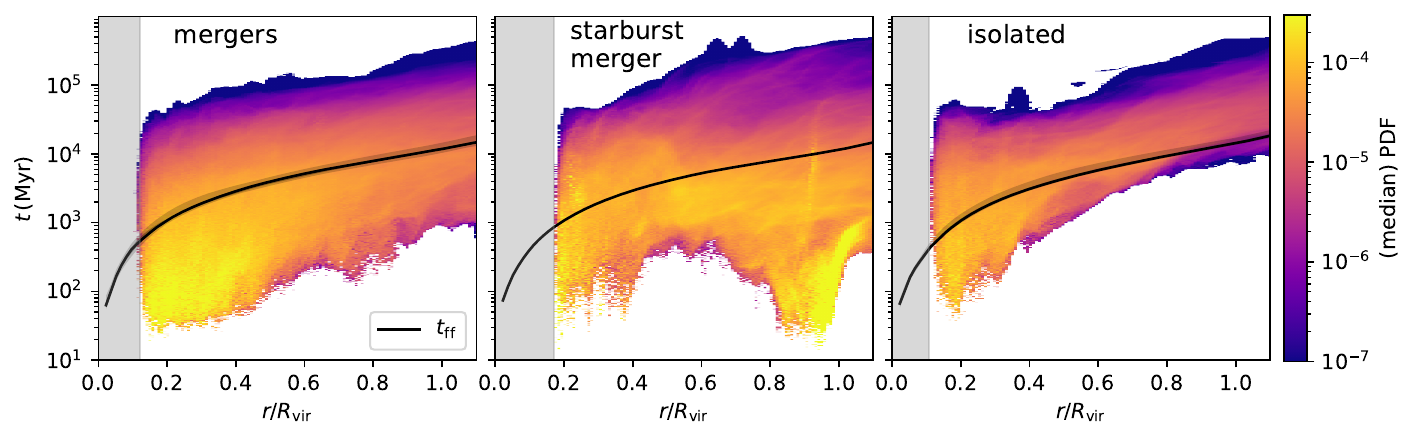}
    \caption{Radial profiles of the median density-weighted cooling time (the median value is computed using the same method as in \figref{fig:cgm_phase}) in mergers (left), the starburst merger (middle), and the isolated galaxies (right), superposed with the median free-fall time (black solid line) and corresponding 16th to 84th percentile ranges of each category in light grey shaded regions. The 16th to 84th percentiles of the median free-fall time are shown by the dark grey shaded region, which is, however, very close to the median value and thus almost invisible in the figure. In isolated galaxies, $t_\mathrm{cool}$ generally exceeds $t_\mathrm{ff}$ throughout most of the halo except in the inner region, consistent with hot-mode accretion. In contrast, merging systems exhibit significant regions with $t_\mathrm{cool} < t_\mathrm{ff}$, especially in the CGM, indicating a transition to cold-mode accretion where radiative cooling operates efficiently. Note that the short cooling times in the outer regions ($r \gtrsim 0.8 R_\mathrm{vir}$) of the starburst merger reflect the accretion of warm, dense gas from the surrounding large-scale structure rather than merger-driven cooling within the established CGM.}
    \label{fig:TcTff}
\end{figure*}

The enhanced cool gas inflows observed in merging galaxies can be attributed to more efficient radiative cooling in their CGM. To quantify this effect, we examine the ratio of cooling time to gravitational free-fall time, which is a key diagnostic of the dominant gas accretion mode. We compute the density-weighted cooling time $t_\mathrm{cool} = 3kT/(2n\Lambda)$ as a function of galactocentric radius for the three categories, where $\Lambda$ is the radiative cooling coefficient. For comparison, we calculate the gravitational free-fall time $t_\mathrm{ff} = \sqrt{2r/|\nabla\Phi|}$, where $\Phi$ is the gravitational potential. Systems with $t_\mathrm{cool} < t_\mathrm{ff}$ are expected to exhibit cold-mode accretion, in which gas cools radiatively before falling onto the central galaxy, whereas those with $t_\mathrm{cool} > t_\mathrm{ff}$ experience hot-mode accretion, in which gas remains at elevated temperatures during infall and slowly cools near the galaxy after settling into the central regions. We present the results in \figref{fig:TcTff}.

In the isolated control galaxies, $t_\mathrm{cool}$ substantially exceeds $t_\mathrm{ff}$ throughout the CGM except in the inner region, consistent with a predominantly hot accretion mode in which inflowing gas remains at temperatures near the virial temperature and cools inefficiently. In contrast, merging systems including the starburst merger exhibit $t_\mathrm{cool} < t_\mathrm{ff}$ over significant portions of the CGM, extending to approximately half the virial radius, indicating a transition to cold-mode accretion wherein gas undergoes rapid radiative cooling before reaching the central galaxy. This condition is particularly pronounced in the starburst merger, where $t_\mathrm{cool} < t_\mathrm{ff}$ holds nearly everywhere within the halo at $r \lesssim 0.8 R_\mathrm{vir}$. We note that the anomalously short cooling times observed in the outermost regions of the starburst merger ($r \gtrsim 0.8 R_\mathrm{vir}$) arise from the presence of warm, high-density gas accreting from the surrounding large-scale structure and several galaxies near the edge of the halo rather than from merger-induced cooling processes within the established CGM itself. 

These results demonstrate that galaxy interactions significantly alter the thermal structure of the CGM, triggering a transition from hot-mode to cold-mode accretion. In isolated galaxies at this mass scale, the CGM is expected to be approximately single phase, with a characteristic temperature near the virial temperature of $\sim10^5\unit{K}$ and density of $\sim 10^{-5}\,\mathrm{cm^{-3}}$ such that its radiative cooling operates on timescales longer than the dynamical time, allowing gas to remain near the virial temperature as it accretes. However, mergers disrupt this equilibrium through multiple physical mechanisms: tidal compression and shocks heat and compress infalling streams and CGM gas, temporarily raising densities while the post-shock cooling drives gas toward the thermally unstable regime where $t_\mathrm{cool} \sim t_\mathrm{ff}$. We suspect that stripped cool gas from satellites is also physically relevant because it can seed turbulent mixing layers with the ambient hot CGM, promoting mixing-induced cooling \citep{Ji2019}. The subsequent fragmentation and condensation of this thermally unstable gas creates a multiphase medium with substantial cool, dense clumps embedded in the hot halo, as evidenced by the extended tail toward lower temperatures in the phase diagrams (\figref{fig:cgm_phase}). Once $t_\mathrm{cool} < t_\mathrm{ff}$ is established over significant halo volumes, gas rapidly loses thermal support and falls toward the galaxy before pressure forces can redistribute it, establishing a self-sustaining cold accretion mode. This transition explains the enhanced cool gas inflows (\figref{fig:CGM_flux}) and provides additional fuel for the time-averaged sSFR enhancement in merging systems.

\subsection{Gas redistribution in the CGM}

\begin{figure*}
    \centering
    \subfloat[Mergers (include the starbusrt merger)\label{fig:merger_sankey}]{
        \includegraphics{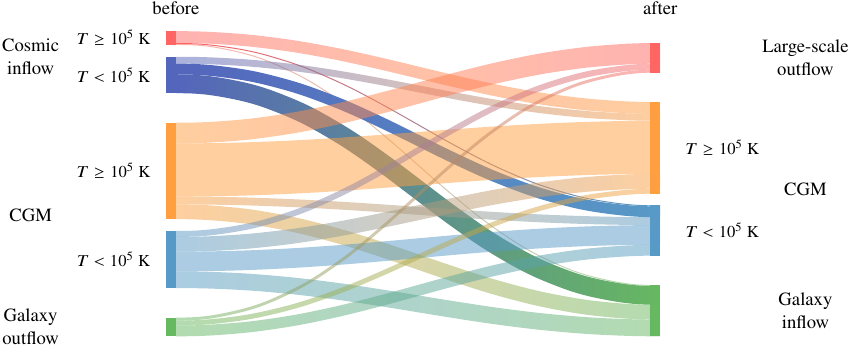}
    }
    \\
    \subfloat[Isolated galaxies\label{fig:isolated_sankey}]{
        \includegraphics{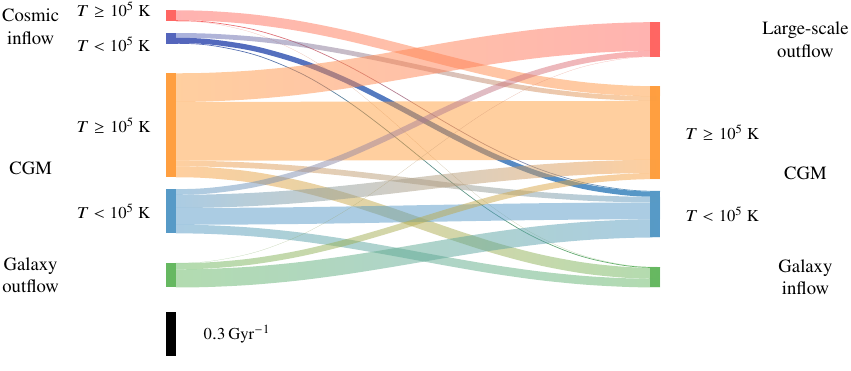}
    }
    \caption{Sankey diagrams of the averaged specific mass transfer rate (sMTR; \myeqref{eq:flow_rate}) for (a) merging galaxies \edit{(include the starburst merger) and (b)} isolated galaxies. Left (right) nodes show the total sMTR from (into) each component at the earlier (later) redshift. Components: galaxy inflow (gas accreted into the ISM or forming stars), galaxy outflow, cold/cool CGM ($T<10^5\unit{K}$), warm/hot CGM ($T\geq10^5\unit{K}$), cold/cool and warm/hot cosmic inflow, and large-scale outflow. Flow widths scale with sMTR, and the absolute scale in $\unit{Gyr^{-1}}$ is shown in the bottom-left corner. Solid, fully opaque pathways mark the significantly enhanced channels from cosmic inflow to cold/cool CGM and the galaxy. Overall, merging systems exhibit enhanced sMTR across nearly all pathways, and the effective transfer rate from cold/cool cosmic inflow to galaxy inflow is amplified by factors of $\sim 30$ compared to isolated galaxies, reflecting rapid CGM processing and efficient fueling of the ISM.}
    \label{fig:sankey_diagram}
\end{figure*}

To quantify the intensity of mass exchange between baryon reservoirs in and around the CGM, we introduce the specific mass transfer rate (sMTR), defined analogously to the specific star formation rate (sSFR) but normalized by the CGM mass of each halo:
\begin{equation}
\mathrm{sMTR} \equiv \frac{\dot{M}}{M_\mathrm{CGM}},
\label{eq:flow_rate}
\end{equation}
where $\dot{M}$ is the mass transfer rate and $M_\mathrm{CGM}$ is the total CGM mass of the corresponding halo. The term sMTR is chosen to mimic sSFR: instead of normalizing the star formation rate by the stellar mass, we normalize each mass transfer rate by the CGM mass of the same halo. Its physical meaning is equivalent to $t_\mathrm{cycle}^{-1}$, where $t_\mathrm{cycle}\equiv M_\mathrm{CGM}/\dot{M}$ is the characteristic time needed for this pathway to transfer one CGM-mass equivalent.

To evaluate pathway-specific sMTR, we track all halo baryons by their particle IDs over matched time intervals (from companion galaxy entry into the halo to coalescence; identical elapsed time for isolated controls), dividing them into galaxy inflow/outflow, CGM, and large-scale inflow/outflow. Here ``galaxy inflow'' encompasses all gas entering the galaxy (both central galaxies and satellites included) that subsequently joins the ISM or forms stars. Similarly, ``galaxy outflow'' includes gas leaving any galaxy in the halo, so this channel includes stripped material from satellites and merging companions as well as feedback-driven outflows. The CGM and cosmic inflow are further split into cold/cool ($T<10^5\unit{K}$) and warm/hot ($T\geq10^5\unit{K}$) phases. We compute the mass transfer rate between each component pair at successive times and present the sample-averaged results as Sankey diagrams\footnote{We note that by this method the diagram will not show gas recycling. However, since we are focusing on the net redistribution among gas reservoirs, it would not affect the interpretation of our results.} in \figref{fig:sankey_diagram}.

\figref{fig:sankey_diagram} illustrates two key features of merger-driven CGM cycling:

First, mergers drive a system-wide enhancement of multiphase gas cycling across the CGM, strengthening exchange pathways between cosmic inflow, CGM phases, galaxy inflow, and large-scale outflow. The sMTR along each pathway is proportional to the thickness of the corresponding flow in \figref{fig:sankey_diagram}, and the thickness of nearly all inter-component flows is substantially increased in mergers relative to isolated galaxies.

Second, among all enhanced pathways, the most dramatic increase occurs along the cold/cool cosmic inflow $\to$ cold/cool CGM and galaxy channel, which is highlighted by the solid, fully opaque flow in \figref{fig:sankey_diagram}. In particular, the sMTR of cold/cool cosmic inflow $\to$ galaxy is amplified by factors of $\sim27$ in mergers relative to isolated galaxies. In terms of the characteristic cycling timescale, this corresponds to a reduction from $t_\mathrm{cycle}\sim10^2\unit{Gyr}$ in isolated systems to $\sim10\unit{Gyr}$ in mergers, i.e., without merger-driven perturbations, the cold/cool cosmic inflow would take many Hubble times before accreting onto the galaxy, whereas mergers compress this timescale by more than an order of magnitude, rapidly delivering fresh cosmological material to the ISM and fueling star formation through fast-cycling CGM. This is consistent with the enhanced cool gas inflows observed in \figref{fig:CGM_flux} and the transition to cold-mode accretion in \figref{fig:TcTff}.

We also note that the sMTR of the galaxy-outflow contribution to the cold/cool CGM is $\sim0.06\unit{Gyr^{-1}}$, smaller than the cooling-related contribution from warm/hot CGM to cold/cool CGM and galaxy inflow of $\sim0.12\unit{Gyr^{-1}}$. This comparison does not include CGM-processed cosmological inflow, which is an additional non-stripping source of cold/cool CGM gas. Therefore, stripped satellite material alone cannot explain the enhanced cold/cool CGM, although it may still be highly relevant by seeding mixing-induced cooling.

Despite this enhanced accretion \edit{from CGM to galaxies} in merger systems, the CGM is not depleted: continuous replenishment from cosmic inflow and galaxy outflows sustains a comparable total CGM mass budget (as shown by \figref{fig:compare_density_and_T}), while cold/cool cosmic inflow becomes the dominant accretion channel, overtaking the warm/hot mode that prevails in isolated systems. To sum up, the CGM in mergers is thus distinguished not by its mass but by its dynamics: it cycles gas far more rapidly, efficiently supplying fresh material to the ISM and fueling star formation on dynamical rather than cosmological timescales.

\subsection{Ion column densities}\label{sec:ion}

\begin{figure*}
    \centering
    \subfloat[Column densities in merging and isolated galaxies\label{fig:column_density}]{
    \includegraphics[width=0.95\linewidth]{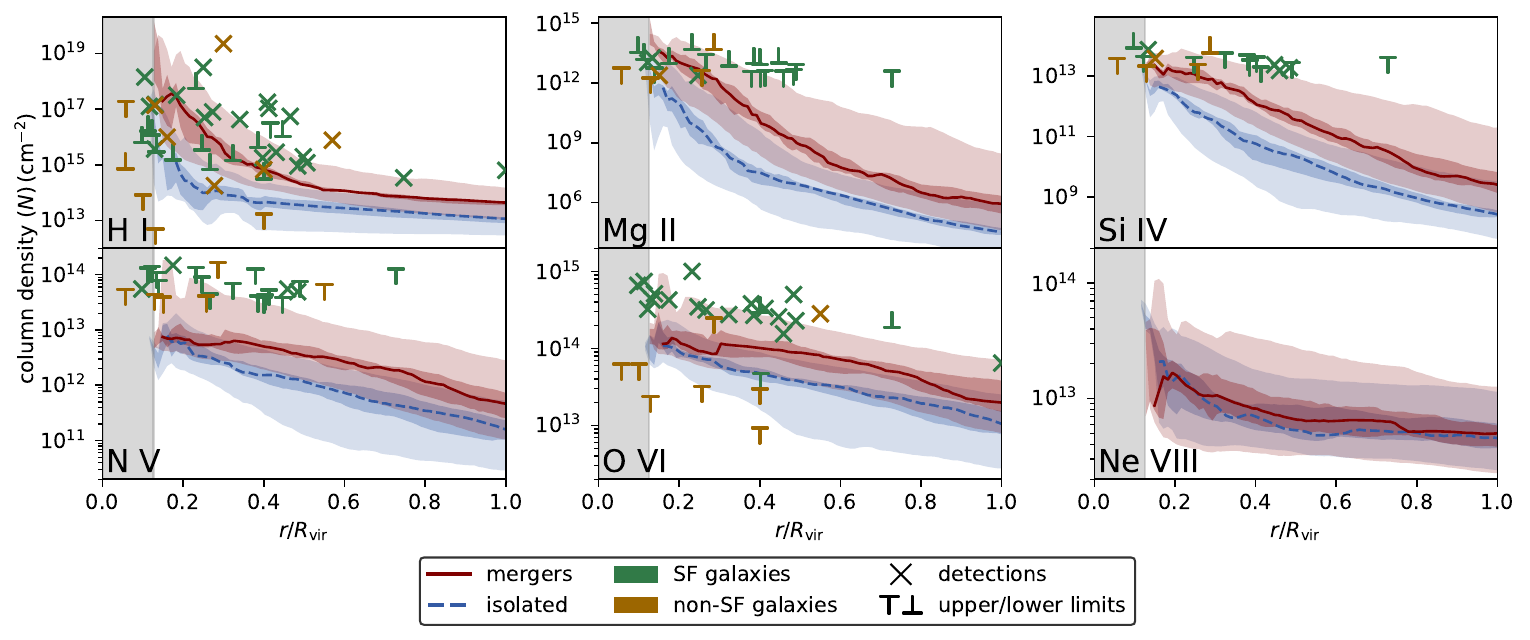}
    }
    \\
    \subfloat[Ratios of column densities between merging and isolated galaxies\label{fig:column_density_ratio}]{
    \includegraphics[width=0.95\linewidth]{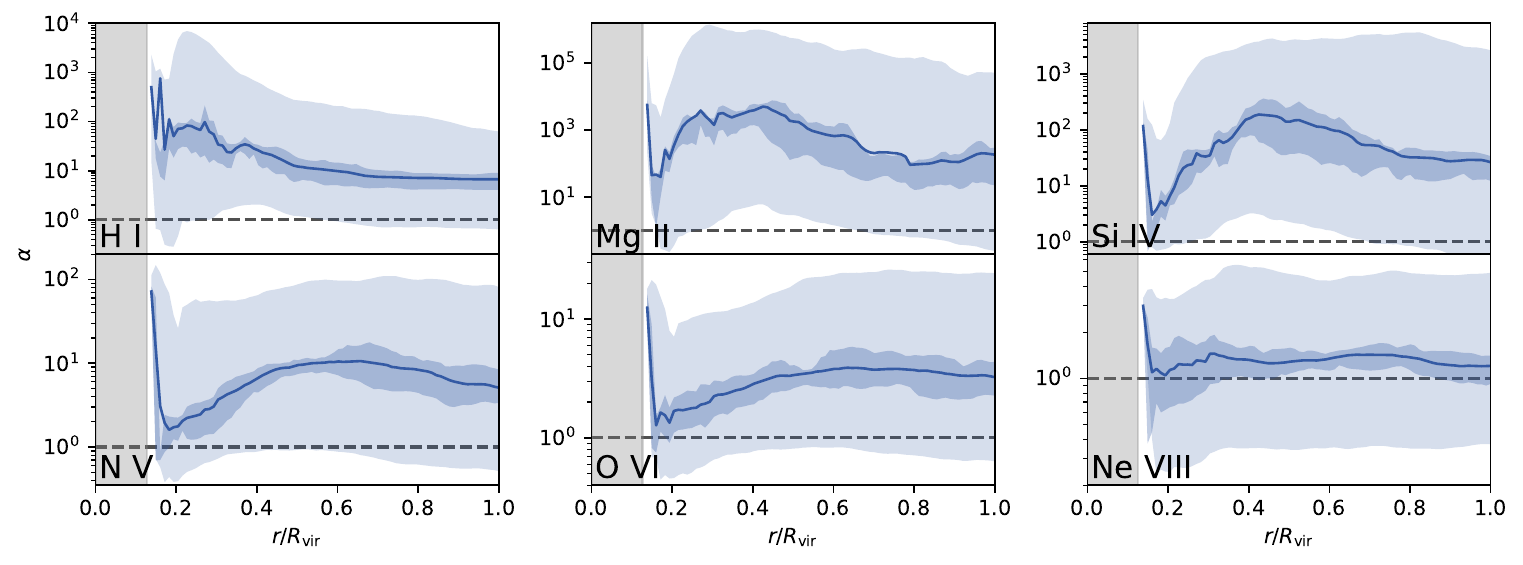}
    }
    \caption{(a): Radial profiles of ion column density for merging galaxies (red solid line) and isolated control galaxies (blue dashed line) as a function of galactocentric radius. Different panels show different ions, as labeled in the bottom left corner. Lines indicate median values, and inner darker shaded regions show the 16th to 84th percentiles of the median values via bootstrapping. Outer lighter region indicates the 16th to 84th percentiles of the category. Observational data from \cite{Werk13,Prochaska2017b,Prochaska2017a,Johnson15,Burchett2019} are shown for comparison, with crosses representing detections and T symbols denoting upper/lower limits for star-forming (dark green) and non-star-forming (dark yellow) galaxies. The observational samples are selected to match the stellar mass ($10^{10}<M_\star<10^{11}\Msun$) of our simulated galaxies. (b): Ratios of ion column densities between merging and isolated galaxies (denoted as $\alpha$). Ratios are computed for each of the 44 matched merger-control pairs constructed from the 18 merging galaxies and 20 control galaxies, before deriving the median and percentile ranges. As in \figref{fig:column_density}, the solid lines show the median values, the darker shaded regions indicate the 16th–84th percentile confidence interval of the median estimated by bootstrapping, and the lighter shaded regions show the 16th–84th percentile range of the full pair distribution. The horizontal dashed grey line at y=1 marks equal column densities in merging and isolated galaxies.}
    \label{fig:all_column_density}
\end{figure*}

We now examine the observational signatures of the enhanced multiphase CGM in merging galaxies by computing ion column densities along sightlines through the simulated halos. We select \ion{H}{1}, \ion{Mg}{2}, \ion{Si}{4}, \ion{N}{5}, \ion{O}{6}, and \ion{Ne}{8}, which trace cold gas, $\approx 10^4\unit{K}$ gas, $\approx 10^{4.9}\unit{K}$ gas, $\approx 10^{5.2}\unit{K}$ gas, $\approx 10^{5.5}\unit{K}$ gas and $\approx 10^{5.8}\unit{K}$ gas, respectively. \figref{fig:column_density} presents the radial profiles of column density for the six key ions in both merging (red) and isolated (blue) systems, compared with available observational measurements.\footnote{We note that while we analyze our simulations at $z = 0$, the observational samples span $0 < z < 0.4$ to ensure sufficient statistics, and narrowing to $z \approx 0$ would yield too few observations for meaningful comparison. Some discrepancies between simulations and observations may therefore arise from this redshift mismatch.} Following \cite{Ji2020}, we use {\small Trident} \citep{hummels2017trident} to calculate these column densities. \edit{The UV background from \cite{FG09} (FG09 hereafter) reproduces observation of low ions better than the UV background from \cite{HM12} (HM12 hereafter), but is less accurate in high ions than HM12 since HM12 treats the contribution of AGN in a more sophsticated way. Thus We use FG09 to predict the column density of \ion{H}{1}, \ion{Mg}{2}, and \ion{Si}{4}, and HM12 for \ion{N}{5}, \ion{O}{6}, and \ion{Ne}{8}}. To conveniently compare the column densities of mergers and isolated galaxies, and emphasize differences between them, we also plot the ratio of column density between mergers and isolated galaxies, $N_\mathrm{merger}/N_\mathrm{isolated}$ (denoted as $\alpha$ in the figure), in \figref{fig:column_density_ratio}. We calculate the ratio pair by pair and then apply the scatter calculation method introduced in \S~\ref{sec:scatter_calculation}. The primary result is that merging galaxies exhibit statistically elevated column densities for low-ionization and intermediate-ionization species relative to isolated systems, while high-ionization tracers show minimal differences between the two populations.

Merging systems display substantial enhancements in \ion{H}{1}, \ion{Mg}{2}, and \ion{Si}{4} column densities. \ion{H}{1} shows an excess of factors $\sim 10$--$100$ in the inner to intermediate CGM ($0.2 R_\mathrm{vir} \lesssim r \lesssim 0.4 R_\mathrm{vir}$), declining to $\sim 0.5$ dex at larger radii. This result is similar to the observational work presented by \cite{Wang2025}, who utilize \ion{H}{1} surface density images of 35 galaxies from the FEASTS program and find that in strongly tidal-interacting galaxy systems the \ion{H}{1} profiles tend to deviate from other galaxies and become more flattened and thus have higher \ion{H}{1} column density at larger radii than other galaxies.

Similarly, \ion{Mg}{2} exhibits an enhancement of approximately one order of magnitude at $R \sim 0.2 R_\mathrm{vir}$, increasing to larger factors beyond $0.4 R_\mathrm{vir}$. \ion{Si}{4} displays a comparable pattern, with mergers showing column densities exceeding those of isolated galaxies by more than one dex at $R > 0.2 R_\mathrm{vir}$. These enhancements reflect the increased cool gas content in merging systems demonstrated in \figref{fig:cgm_phase}, as these ions trace gas at temperatures $T \sim 10^4$--$10^5\unit{K}$. Compared with observational data, the simulated column densities for \ion{Mg}{2} and \ion{Si}{4} in merging galaxies are not inconsistent with available measurements: the upper envelope of our simulated merger distributions traces the observed upper limits, though most observational constraints are upper or lower limits rather than detections, making definitive comparisons difficult. The simulated \ion{H}{1} columns are slightly lower than observed values, but merging systems show better agreement than isolated galaxies. By excluding satellite galaxies, our sample may underestimate gas-rich satellite contributions to observed \ion{H}{1} absorption.

Crucially, this excess in column density of \ion{H}{1}, \ion{Mg}{2}, and \ion{Si}{4} does not result from higher overall gas densities, as the total CGM gas density profiles are comparable between merging and isolated galaxies (shown in \figref{fig:compare_density_and_T}). It is also not the result of higher metallicities in mergers: we have compared the metallicities of merging and isolated galaxies, and found that mergers only show a factor of $\lesssim 3$ enhancement of metallicity in the inner halo, and even a slight decrease in the outer halo. These metallicity differences cannot explain the large enhancements in column densities. Instead, the elevated ion columns arise directly from the altered thermal phase structure of the CGM: mergers substantially increase the mass fraction of cool gas (as shown in the extended tail toward lower temperatures in \figref{fig:cgm_phase}), thereby boosting the abundance of low- and intermediate-ionization species without significantly changing the total baryon content.

For intermediate-temperature ions, \ion{N}{5} shows modest differences, with merging galaxies exhibiting relatively flat column density profiles between $0.1 R_\mathrm{vir}$ and $0.4 R_\mathrm{vir}$ while isolated systems decline steadily, resulting in an excess of $\gtrsim 0.5$ dex at $r \sim 0.4 R_\mathrm{vir}$. \ion{O}{6}, which traces gas near $T \sim 10^{5.5}\unit{K}$, displays even smaller differences: merging galaxies show only a $\lesssim 0.5$ dex enhancement at all radii compared to isolated systems. This modest excess is consistent with the relatively minor changes in warm gas content between the two populations. While simulated mergers exhibit higher \ion{N}{5} and \ion{O}{6} column densities than isolated galaxies, both populations appear somewhat underestimated compared to observations, with observational values typically lying near the upper envelope of our merger distribution and exceeding simulation medians by factors of $\sim 1$. This discrepancy may reflect limitations in resolving turbulent mixing layers at a few $10$ pc scales where \ion{O}{6} is efficiently produced \citep{Begelman1990,Ji2019}, the potential contribution of photoionized \ion{O}{6} supported by cosmic ray pressure \citep{Ji2020}, or the lack of AGN feedback physics, which may enhance \ion{O}{6} column density \citep{Mathews2017,Nelson2018}, none of which are fully captured in our simulations. Also, although gas-rich satellites are present in our simulations, we exclude them from the ion column density analysis here, which may lead to underestimation of the column density of certain species.

In contrast, \ion{Ne}{8}, a tracer of hot gas ($T \gtrsim 10^6\unit{K}$), exhibits negligible differences between merging and isolated galaxies. Both populations display similar \ion{Ne}{8} column densities declining from $N_{\mathrm{Ne\,VIII}} \sim 5 \times 10^{13}\unit{cm^{-2}}$ at $r \sim 0.1 R_\mathrm{vir}$ to $\sim 6 \times 10^{12}\unit{cm^{-2}}$ near $R_\mathrm{vir}$. This lack of enhancement reflects the comparable hot gas fractions in both samples, as hot gas remains a minor component in both merging and isolated systems for the given halo mass range. Observational measurements of \ion{Ne}{8} are not included in the figure due to the lack of \ion{Ne}{8} observations within the mass and redshift ranges ($0<z<0.4$), while as indicated in \cite{Wijers2024}, the FIRE-2 model is consistent with the observed \ion{Ne}{8} column density in $\sim10^{12}\Msun$ halos but may under-predict its scatter.

\section{Discussion and Conclusions} \label{sec:discussion}

\subsection{Discussion}

We present a systematic comparison of circumgalactic medium (CGM) properties in merging versus isolated galaxies using the cosmological FIREbox simulation \citep{Feldmann2023}. By analyzing matched samples of 18 merging and 20 isolated systems with stellar masses $M_\star \sim 10^{10}$--$10^{11} \, \mathrm{M}_{\odot}$ at $z=0$, we quantify how galaxy interactions fundamentally alter CGM thermodynamics, gas flows, and observational signatures.

Our results reveal how galaxy interactions reshape the circumgalactic medium and regulate baryon flows. Mergers alter the thermal structure of intermediate-mass halos by enabling efficient radiative cooling throughout significant portions of the CGM, overcoming the relatively long cooling times characteristic of isolated systems at similar masses. Gravitational interactions introduce compressive motions, shocks, and mixing that reduce cooling times below free-fall times ($t_\mathrm{cool} \lesssim t_\mathrm{ff}$), producing the enhanced multiphase structure and elevated cool gas content we observe. In the context of mergers, the CGM emerges not as a passive reservoir but as a dynamically responsive system: merger-driven perturbations amplify both inflows and outflows across all temperature phases, particularly for the cool phase and $\sim10^6\unit{K}$ gas by a factor of $\gtrsim 10$. This sensitivity implies that time-averaged models of CGM properties may systematically underestimate the true variability driven by interactions in realistic cosmological environments.

Our work emphasizes the critical role of the CGM as a rapidly cycling reservoir during mergers. Gas cycling analysis (\figref{fig:sankey_diagram}) reveals that mergers enhance mass transfer rates along nearly all pathways, most strikingly amplifying the effective transfer rate from cold/cool cosmic inflow to galaxy inflow by factors of \edit{$\sim27$}, compared to isolated systems. This reflects rapid CGM processing: merger-driven perturbations compress the inflow$\to$CGM$\to$galaxy sequence by reducing cooling times below free-fall times, enabling inflowing gas to be cycled through CGM phases and accreted on dramatically shortened timescales, rather than cooling quiescently in the CGM as in isolated systems. Despite this acceleration, the total CGM mass remains comparable to isolated systems, underscoring the CGM's role as a dynamic reservoir whose mass transfer rate --- not mass --- governs galaxy fueling during mergers. This cosmological context, including continuous baryon exchange while retaining material within the virial radius, has been emphasized in previous studies \citep{Moster2011,Hani2018,Sparre2022} and cannot be captured by idealized merger simulations that lack ongoing environmental interactions.

The CGM transformation we identify has direct implications for star formation evolution during and after mergers. We observe a systematic enhancement in time-averaged sSFR in our merger sample, and our sample includes one merger-triggered starburst with specific star formation rates approximately two orders of magnitude above typical mergers, consistent with extensive observational evidence for merger-enhanced star formation \citep[e.g.,][]{Ellison2008,Patton2013,Huang2025}. This comparison indicates that the CGM accretion-mode transition during merging events provides additional fuel for star formation. However, the CGM also plays a central role in galaxy quenching: post-merger galaxies are significantly more likely to shut down star formation than isolated systems \citep{Ellison2022,Li2023}. We suggest that both outcomes, including both starbursts and quenching, are linked to the same CGM processes operating at different merger phases. The enhanced radiative cooling and amplified inflows that fuel enhanced star formation during active merging can also be disrupted by powerful feedback or gas exhaustion in later merger stages, potentially contributing to subsequent quenching. While AGN feedback is often invoked as a quenching mechanism \citep{Hopkins2008,Pontzen2017,Davies2022,Wellons2023,Byrne2024}, the detailed pathways remain debated \citep{RodriguezMontero2019,Quai2023}, particularly for the intermediate-mass systems studied here.

Our simulations predict observational signatures that can be tested with CGM surveys. We find elevated ion column densities for low- and intermediate-ionization species in merging systems, reflecting the enhanced cool gas content. Although simulated column densities are slightly lower than observed values, which is likely due to limited resolution of sub-kpc turbulent mixing layers \citep{Ji2019,Fielding2020,Tan2021}, lack of AGN physics \citep{Mathews2017,Nelson2018}, and exclusion of satellite galaxies, observations of mergers within our simulated mass range are still sparse. Observational validation faces two challenges: most CGM absorption surveys do not distinguish between isolated and merging galaxies, while many merger studies focus on galactic properties rather than the CGM \citep[e.g.,][]{Ellison2008,Patton2013,Liu2011,Jin2021}. Notable exceptions include \citet{Dutta2020} who found that galaxy groups exhibit significantly richer \ion{Mg}{2} content and more complex kinematics than isolated galaxies, with similar enhancements reported by \cite{Nielsen2018} and \cite{Huang2021}. Complementary 21~cm emission observations \citep{Westmeier2022,Wang2023,Lin2025} can directly map neutral hydrogen distributions, with \cite{Wang2023} demonstrating that galaxy interactions trigger IGM cooling and generate diffuse \ion{H}{1}, which is an in-situ gas generation mechanism via enhanced radiative cooling \citep{Lin2025} consistent with our findings.

\subsection{Caveats}

A few caveats should be kept in mind when interpreting our results.

\emph{Sample size:} Our sample comprises 18 merging galaxies and 20 isolated control galaxies, which represents a relatively modest statistical sample. This limited size may not fully capture the full diversity of merger configurations (e.g., mass ratios, orbital parameters, merger stages) and their varied effects on CGM properties. Additionally, the inherent stochasticity in merger dynamics and CGM evolution means that larger samples are needed to robustly quantify systematic trends and reduce uncertainties. Future studies leveraging larger cosmological simulation volumes or extensive observational surveys would help validate and extend our findings across a broader parameter space.
    
\emph{Numerical resolution:} We analyze the FIREbox simulation, where the adaptive spatial resolution varies with local gas density, with a baryonic mass resolution of $6.26 \times 10^4 \, \mathrm{M}_{\odot}$. In the CGM, where the typical hydrogen number density of warm/hot gas is around $10^{-4}\unit{cm^{-3}}$, this corresponds to an effective spatial resolution of $\sim 1\unit{kpc}$ or less \citep{FG23}. While sufficient to capture large-scale CGM structures and flows, this resolution may not fully resolve small-scale thermal instabilities \citep{McCourt2018}, turbulent cascades, and mixing processes \citep{Ji2019,Gronke2022} that operate on sub-kpc scales, thus potentially underestimating the efficiency of radiative cooling and the production of multiphase gas \citep{Hummels2019}. This may contribute to the systematically lower ion column densities \citep{Peeples2019,Voort2019} we find compared to observations, particularly for low-ionization species that trace cool, dense gas produced by mixing layers. However, we must also emphasize that although the formation process of cold/cool gas may be under-resolved, the cold/cool, dense gas itself could be better resolved than warmer gas due to the meshless-finite-mass (MFM) method \citep{Hopkins2015} used in FIREbox simulations. Thus higher-resolution simulations will be valuable to further investigate the formation and evolution of cold/cool gas in the CGM and test our findings.
    
\emph{Non-thermal physics:} The FIREbox simulation does not include magnetic fields or cosmic rays, both of which may significantly influence CGM properties. Magnetic fields can provide non-thermal pressure support, modify the growth rates of hydrodynamic instabilities \citep{Ji2018,Wibking2025}, and suppress thermal conduction \citep{Hopkins2020}, thereby altering the structure and survival of cool gas clouds. Cosmic rays, if dynamically important, can drive galactic winds more efficiently \citep{Hopkins2021,Ruszkowski2017}, increase the scale height of the CGM, and modify its thermal and ionization structure by providing additional pressure support and heating through streaming instabilities \citep{Zweibel2013,Bai2019,Weber2025}. These effects could produce a more diffuse and ``fluffy'' CGM, affecting the thermal and ionization state of the CGM, potentially increasing predicted column densities for certain ion species and improving agreement with observations, as shown explicitly in \citep{Butsky2018,Butsky2020,Ji2020}. The absence of these processes in our simulations should be considered when comparing with observational data.
    
\emph{AGN feedback:} The FIREbox simulation does not incorporate AGN feedback, including jet-driven outflows and radiative heating from accretion disks. AGN feedback can strongly affect the evolution of host galaxies \citep{Ho2008,Somerville2015,Su2021,MACER3D}, inject significant amounts of energy and momentum into the CGM \citep{Fabian2012,Heckman2014,Wright2024}, suppress gas cooling and inflow \citep{Dekel2009,Nelson2015,Zhu2023}, and dramatically alter the multiphase structure of the CGM \citep{Suresh2015,Zinger2020,Talbot2021}. However, given that our sample galaxies have relatively modest stellar masses ($M_\star \sim 10^{10}$--$10^{11} \, \mathrm{M}_{\odot}$) and likely host correspondingly lower-mass black holes, AGN feedback may play a less significant role in these systems. Simulation studies of AGN feedback in dwarf galaxies suggest negligible impacts on CGM-scale properties \citep{Su2025}, though the effects remain uncertain for the intermediate mass range explored in this work. We note that several other FIRE simulations include AGN feedback physics \citep[e.g.,][]{Hopkins2023,Wellons2023,Byrne2024,Ponnada2025}, providing an opportunity for future investigations of the role of AGN feedback in galaxy mergers.

\subsection{Conclusions}

Our main conclusions are:

\begin{itemize}
    \item \emph{Mergers drive enhanced multiphase structure and amplified gas flows.} Merging systems display substantially broader distributions in temperature-density phase space, with pronounced enhancements in cool ($T \sim 10^4\unit{K}$), dense gas compared to isolated galaxies. Both inflow and outflow mass fluxes increase across a wide range of temperature phases, and the enhancement is most pronounced in the cool ($\sim 10^4\unit{K}$) and hot ($\sim10^6\unit{K}$) phases by $\gtrsim 1\unit{dex}$, with the median value of cold/cool gas net outflow enhanced by a factor of $\sim15$. This cool gas supply arises predominantly from in-situ radiative cooling within the CGM rather than from pre-existing cold streams. Notably, while the total baryon content within the virial radius remains comparable between merging and isolated systems, mergers fundamentally alter the phase distribution and the multiphase flow dynamics of CGM gas.
    
    \item \emph{Mergers trigger enhanced radiative cooling in the CGM.} Merging galaxies exhibit cooling times shorter than free-fall times ($t_\mathrm{cool} \lesssim t_\mathrm{ff}$) throughout significant portions of their CGM. This condition, largely absent in isolated systems of comparable mass, indicates that mergers induce efficient radiative cooling. The resulting multiphase structure and enhanced cool gas content directly enable the amplified inflow rates described above, linking the cooling physics to the observed mass flux enhancements.
    
    \item \emph{Mergers fundamentally accelerate CGM gas cycling.} The CGM-mass-normalized specific mass transfer rate (sMTR) is enhanced along nearly all pathways in mergers. Most notably, the effective transfer rate from cold/cool cosmic inflow to galaxy inflow is amplified by factors of \edit{$\sim$27 in mergers}, corresponding to a reduction in the characteristic cycling timescale from $t_\mathrm{cycle}\sim10^2\unit{Gyr}$ in isolated systems to $\sim10\unit{Gyr}$ in mergers. This reflects rapid CGM processing: inflowing gas is efficiently cycled through intermediate CGM phases on dramatically shortened timescales, rather than cooling quiescently in the CGM as in isolated systems. The CGM mass budget is maintained by replenishment from cosmic inflow and galaxy outflows; it is the cycling rates, not the reservoir mass, that distinguishes merging from isolated systems.

    \item \emph{Mergers produce elevated low- and mid-ion column densities in the inner CGM.} The enhanced cool gas content in merging systems produces systematically elevated column densities for low- and intermediate-ionization species (\ion{H}{1}, \ion{Mg}{2}, \ion{Si}{4}, \ion{N}{5}, \ion{O}{6}) in the inner to intermediate CGM, while high-ionization tracers like \ion{Ne}{8} show negligible differences between populations.
    
\end{itemize}

Despite these caveats, our study provides a novel and systematic analysis of how galaxy interactions reshape the CGM in a cosmological context. Our findings highlight the critical role of mergers in driving enhanced radiative cooling, multiphase structure, and rapid gas cycling in the CGM, with direct implications for galaxy evolution during these transformative events. Future work with larger samples, higher resolution, and additional physics will further refine our understanding of the complex interplay between galaxy interactions and their circumgalactic environments.

\datastatement{The data supporting the plots within this article are available in a Zenodo repository at \url{doi: 10.5281/zenodo.23033340}. A public version of the GIZMO code is available at \url{http://www.tapir.caltech.edu/~phopkins/Site/GIZMO.html}. FIRE-2 simulations are publicly available \citep{wetzel2023public} at \url{http://flatathub.flatironinstitute.org/fire}. Additional data, including initial conditions and derived data products, are available at \url{https://fire.northwestern.edu/data/}.}

\begin{acknowledgments}
    Authors are supported by the NSF of China (grants 12192223, 12522301, 12133008, and 12361161601), the China Manned Space Program (grants CMS-CSST-2025-A08 and CMS-CSST-2025-A10), the National Key R\&D Program of China No. 2023YFB3002502, and the National SKA Program of China (No. 2025SKA0130100). C.-A. F.-G. was supported by NSF through grants AST-2108230 and AST-2307327; by NASA through grants 80NSSC22K0809, 80NSSC22K1124 and 80NSSC24K1224; by STScI through grant JWST-AR-03252.001-A; and by BSF through grant \#2024262. This work was performed in part at the Aspen Center for Physics, which is supported by National Science Foundation grant PHY-2210452. Numerical calculations were run on the CFFF platform of Fudan University, the supercomputing system in the Supercomputing Center of Wuhan University, and the High Performance Computing Resource in the Core Facility for Advanced Research Computing at Shanghai Astronomical Observatory.
\end{acknowledgments}

\software{{\small GIZMO} \citep{Hopkins2015},
          {\small Trident} \citep{hummels2017trident},
          {\small Cloudy} \citep{Cloudy},
          {\small Matplotlib} \citep{hunter2007matplotlib},
          {\small NumPy} \citep{harris2020array},
          {\small SciPy} \citep{virtanen2020scipy},
          {\small yt} \citep{turk2010yt,turk2025introducing}
}
          

\bibliography{ref}
\bibliographystyle{aasjournalv7}

\end{document}